\documentclass[aps,prb,twocolumn,superscriptaddress,showpacs]{revtex4-1}
\usepackage{amsmath,amssymb,graphicx,stmaryrd}

\usepackage{ifthen}
\newboolean{PRBVERSION}
\setboolean{PRBVERSION}{false}

\ifthenelse {\boolean{PRBVERSION}}
{
\newcommand{\rc}{Rapid Communication}
\newcommand{\co}{(color online)\ }

\newcommand{\centred}{centered}

\newcommand{\neighbour}{neighbor}
\newcommand{\behaviour}{behavior}
\newcommand{\favour}{favor}
\newcommand{\Acknowledgements}{Acknowledgments}
\newcommand{\doivpy}[2]{#2}
\newcommand{\arxiv}[2]{arxiv:#1 #2}
\newcommand{\isbn}[1]{}
}
{
\pdfoutput=1
\usepackage{color}
\definecolor{LinkColor}{rgb}{0.256,0.439,0.588}
\usepackage{hyperref}
\hypersetup{
   pdfauthor={X Zhang, J Xu, KSD Beach},
   pdftitle={Frustrating antiferromagnetic exchange interactions enhance specific valence-bond-pair motifs},
   pdfsubject={Frustrated quantum magnetism},
   pdfkeywords={quantum spin model,} {frustration,} {resonating valence bonds,} {spin liquid},
   colorlinks=true,
   citecolor=LinkColor,
   linkcolor=LinkColor,
   urlcolor=LinkColor
   }
\usepackage{mathptmx}
\newcommand{\rc}{paper}
\newcommand{\co}{}

\newcommand{\centred}{centred}

\newcommand{\neighbour}{neighbour}
\newcommand{\behaviour}{behaviour}
\newcommand{\favour}{favour}
\newcommand{\Acknowledgements}{Acknowledgements}
\newcommand{\doivpy}[2]{\href{http://dx.doi.org/#1}{#2}}
\newcommand{\arxiv}[2]{\href{http://arxiv.org/abs/#1}{arxiv:#1 #2}}
\newcommand{\isbn}[1]{ISBN #1}
}

\newcommand{\bra}[1]{\langle #1 \rvert}
\newcommand{\ket}[1]{\lvert #1 \rangle}
\newcommand{\expectation}[1]{\langle #1\rangle}
\newcommand{\overlap}[2]{\langle #1 | #2 \rangle}
\newcommand{\matrixelem}[3]{\langle #1 | #2 | #3 \rangle}

\begin{document}

\title{Frustrating antiferromagnetic exchange interactions enhance specific valence-bond-pair motifs}

\author{Xiaoming Zhang}
\affiliation{Department of Physics and Astronomy, University of Western Ontario, London, Ontario, Canada N6A 3K7 }
\affiliation{Department of Earth Sciences, University of Western Ontario, London, Ontario, Canada N6A 5B7 }

\author{Jin Xu}
\affiliation{Department of Physics, University of Alberta, Edmonton, Alberta, Canada T6G 2E1}

\author{K. S. D. Beach}
\email[Electronic mail:\ ]{kbeach@ualberta.ca}
\affiliation{Department of Physics, University of Alberta, Edmonton, Alberta, Canada T6G 2E1}

\date{October 22, 2013}

\begin{abstract}
We present variational results for the ground state of the antiferromagnetic quantum Heisenberg model with frustrating
next-nearest-{\neighbour} interactions. The trial wave functions employed are of resonating-valence-bond type, elaborated 
to account for various geometric motifs of adjacent bond pairs. The calculation is specialized to a square-lattice cluster 
consisting of just sixteen sites, large enough that the system can accommodate nontrivial singlet dimer correlations
but small enough that exhaustive enumeration of states in the total spin zero sector is still feasible. 
A symbolic computation approach allows us to generate an algebraic expression for the expectation value of 
any observable and hence to carry out the energy optimization exactly. While we have no measurements
that could unambiguously identify a spin liquid state in the controversial region at intermediate frustration,
we can say that the bond-bond correlation factors that emerge do not appear to be consistent with the existence 
of a columnar valence bond crystal. Furthermore, our results suggest that the magnetically disordered region may
accommodate two distinct phases.

\end{abstract}

\ifthenelse {\boolean{PRBVERSION}}
{}
{}

\maketitle

Frustration\cite{Diep05,Lacroix11,Balents10} is believed to be a key ingredient for stabilizing  
magnetically disordered states\cite{Affleck87,Affleck88,Read89} in quantum spin systems.
One of the canonical and most widely discussed frustrated models is the so-called 
$J_1$--$J_2$, a spin-half quantum Heisenberg hamiltonian with 
nearest- and next-nearest-{\neighbour} antiferromagnetic exchange interactions.\cite{Chandra88} 
When defined on a bipartite lattice, the model possesses a weakly frustrated limit, in the vicinity of
$J_2/J_1 = 0$, in which the ground state has a well-defined Marshall sign structure\cite{Marshall55,Richter94,Beach09,Richter10,Zhang13}
and a tendency toward antiferromagnetic order (scrambled by quantum fluctuations in one dimension\cite{Mermin66,Hohenberg67} 
but generally robust for higher-dimensional lattices\cite{Reger89,Weihond91,Beach07a}).
As the relative coupling is tuned up from zero, the effects of the frustration become increasingly disruptive,
to the point where they induce a quantum phase transition.

The $J_1$--$J_2$ model on the linear chain has been extensively
studied, and its {\behaviour} over the full range of relative couplings is known.\cite{White96} 
A crucial anchor for our understanding 
is the famous Majumdar-Ghosh point,\cite{Majumdar69} at $J_2/J_1 = 1/2$, where the ground state is a 
perfect spin-Peierls product state,\cite{Frahm97} i.e., a fluctuationless crystal of dimerized singlet pairs 
or \emph{valence bonds}.\cite{Rumer32,Pauling33,Hulthen38}
What the one-dimensional model clearly demonstrates is that frustration can cause a 
valence bond state that otherwise resonates\cite{Anderson73,Fazekas74,Sutherland88,Liang88}
to freeze into a static bond pattern that breaks translational symmetry.

In two dimensions, however, there is no comparable, exactly solvable point
and no compelling reason to believe that bond crystallization must occur.
There do exist models with farther-{\neighbour} interactions for which the columnar bond crystal is a genuine ground state\cite{Bose91}
(and others with staggered and herring bone dimer states\cite{Bose92}),
but it is unclear whether such models are in any sense ``close'' to the $J_1$--$J_2$ in parameter space.
Indeed, Jiang, Yao, and Balents  argue (forcefully, in Ref.~\onlinecite{Jiang11}) that
the columnar valence bond crystal is {\it not} a good candidate for the intermediate
phase of the square-lattice version of the model. Rather, their density matrix renormalization group
calculations suggest a $\mathbb{Z}_2$ spin liquid with robust gaps to the $S=0$ and $S=1$ excitations.
Hu and coworkers have also presented evidence for a liquid state, but a gapless one.\cite{Hu13}
Their approach is to show that a Gutzwiller-projected fermionic trial wavefunction 
subject to additional Lanczos steps produces extremely good variational energies
in the frustrated regime.
We note that these two claims for a liquid state are just the latest salvos in the 
long-running dispute as to whether\cite{Gelfand89,Gelfand90,Singh90,Zhitomirsky96,Leung96,Kotov99, Singh99,Kotov00,Capriotti00,Saarloos00,Takano03,Mambrini06,Murg09,Reuther10,Reuther11,Yu12}
 or not\cite{Figueirido90,Oguchi90,Locher90,Schulz92,Zhong93,Oitma96,Capriotti01,Zhang03,Capriotti03,Captiotti04a,Capriotti04b,Yunoki04,Wang11,Mezzacapo12,Li12,Wang13}
the ground state has broken translational symmetry. 

This {\rc} attempts to address the issue in the following, modest way. We posit that if a bond 
crystal is {\favour}ed over a spin liquid, there should be a signature in the form of explicit correlations between
pairs of valence bonds---presumably, enhanced (suppressed) correlations between pairs whose arrangement is compatible
(incompatible) with the global crystalline order. This is a reasonable assumption, since we know that
an unbiased short-bond-only state has at most powerlaw dimer correlations.\cite{Albuquerque10,Tang11,Ju12}
Our approach is to construct expressive trial wavefunctions that include 
various pair correlation factors and then to look at the {\behaviour} of the optimized variational parameters that emerge at a 
given level of frustration.

\begin{figure}
\begin{center}
\ifthenelse {\boolean{PRBVERSION}}
{\includegraphics{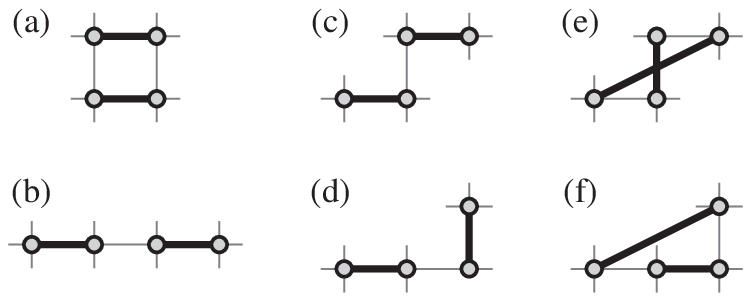}}
{\includegraphics{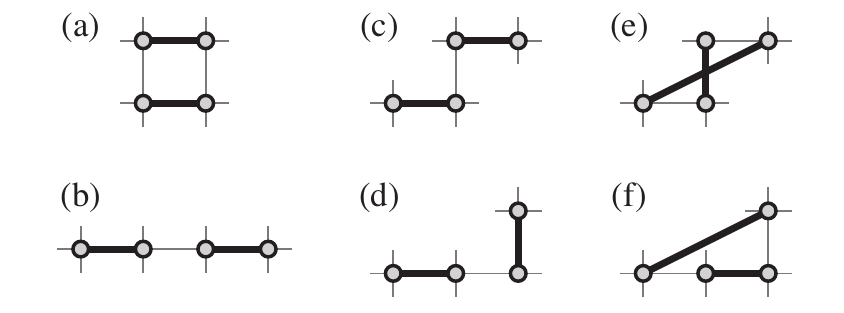}}
\end{center}
\caption{\label{FIG:motifs}
All geometric motifs of two adjacent valence bonds. We give them the monikers
(a) plaquette, (b) line, (c) staggered, (d) herring bone, (e) cross, and (f) ramp.
Up to rotations and reflections, this is the complete set on the $4 \times 4$ lattice,
so long as only bipartite singlet pairings are allowed.
}
\end{figure}

Specifically, we consider a resonating-valence-bond (RVB) wave function of the form
\begin{equation} \label{EQ:rvb_wf}
\ket{\psi(x,y,z)} = \sum_v x^{\alpha(v)}y^{\beta(v)}z^{\gamma(v)} \ket{v}. 
\end{equation}
The summation ranges over all singlet-product states $\ket{v}$ that
are purely bipartite.\cite{Beach06,Beach08} 
On the $4 \times 4$ lattice with periodic boundary conditions---the case to which we specialize---only two kinds of bond
are possible: 
those that are nearest-{\neighbour} [whose end-to-end vector is symmetry equivalent to
$\mathbf{r} = (1,0)$] and those that entangle spins a knight's
move\cite{Iske87} apart [$\mathbf{r} = (2,1)$].

In Eq.~\eqref{EQ:rvb_wf}, we allow for up to three continuous variational parameters, denoted $x$, $y$, and $z$,
each of which appears as a factor raised to an integer power. The exponent $\alpha(v)$ is the number of knight's 
move bonds in $\ket{v}$, whereas
$\beta(v)$ and $\gamma(v)$ count particular motifs of two adjacent bonds, chosen from the six possibilities shown
in Fig.~\ref{FIG:motifs}. The motifs are equivalent under all lattice symmetries and thus
depend only on the relative orientation of the two bonds. The counting is consistent with the convention for
the \emph{correlated} valence bond states established in Ref.~\onlinecite{Lin12}.
We point out that the usual Liang-Doucot-Anderson amplitude product state\cite{Liang88} is recovered from 
Eq.~\eqref{EQ:rvb_wf} with the choice of values $x = h(2,1)/h(1,0)$ and $y=z=1$. In accordance with the standard 
notation, $h(\mathbf{r}_i-\mathbf{r}_j)$ refers to the independent amplitude
for a single bond connecting sites $i$ and $j$ in opposite sublattices.\cite{note}

We argue that, despite its small size, the $4 \times 4$ lattice system provides a good caricature of
the physics of frustration. The lattice is large enough to accommodate all possible pairs of adjacent nearest-{\neighbour} bonds.
Equally important, it accommodates the one crucial long bond that is responsible for the breakdown of the Marshall sign structure:
we know that, initially at least, the knight's move bond is the only one to turn 
negative\cite{Richter94,Beach09,Zhang13} in response to increasing frustration.

{\it Computational approach}. As two of us have pointed out elsewhere,\cite{Zhang13} 
the $4 \times 4$ system can be solved in symbolic form with the help of the Lehmer code.\cite{Lehmer60,Knuth73}
All observables 
\begin{equation}
\expectation{\hat{A}}
= \frac{ \bra{\psi(x,y,z)} \hat{A} \ket{\psi(x,y,z)} }{ \overlap{\psi(x,y,z)}{\psi(x,y,z)} }
 = \frac{A(x,y,z)}{Z(x,y,z)}
\end{equation}
can be computed in the form of a ratio of two mixed polynomials in $x$, $y$, and $z$.
For a given value of the coupling strength $g=J_2/J_1$, the triplet of values $(x,y,z)$ is chosen
to minimize the energy
\begin{equation} \label{EQ:energy_expectation}
E(x,y,z) = \frac{ \bra{\psi} \hat{H} \ket{\psi} }{ \overlap{\psi}{\psi} }
= \frac{ \sum_C x^{\alpha(C)}y^{\beta(C)}z^{\gamma(C)}2^{N_\ell(C)} H(C) }
{  \sum_{C} x^{\alpha(C)}y^{\beta(C)}z^{\gamma(C)}2^{N_\ell(C)} }.
\end{equation}
In the notation of Eq.~\eqref{EQ:energy_expectation}, each configuration is a double dimer covering $C = (v,v')$, and
$N_\ell(C)$ is the number of loops formed by the overlap of states $\langle v \rvert$
and $\ket{v'}$. The exponent $\alpha(C)$ is defined as
$\alpha(C) = \alpha(v) + \alpha(v')$, and there are corresponding definitions for $\beta(C)$ and $\gamma(C)$.
The factors of 2 are loop fugacities that arise from overlaps\cite{Sutherland88} in the overcomplete basis,
and $H(C) = \matrixelem{v}{\hat{H}}{v'}/\overlap{v}{v'}$ is the loop estimator\cite{Beach06} of
the hamiltonian evaluated for configuration $C$.

The energy minimization is carried out by steepest descent (and results verified
afterwards by a hierarchical global search). All that is required is knowledge
of the local downhill direction
\begin{equation} \label{EQ:downhill}
-\nabla \biggl(\frac{E(x,y,z)}{Z(x,y,z)}\biggr) = \frac{-(\nabla E) Z + E \nabla Z}{Z^2},
\end{equation}
where the gradient 
$\nabla = (\partial_x, \partial_y, \partial_z)$ 
is expressed in the variational parameter coordinates.
Since both $E$ and $Z$ have rational polynomial form, 
the downhill vector does too; hence Eq.~\eqref{EQ:downhill} is straightforward to compute, so long as we take
care not to overrun the computer's finite precision.
Our strategy to cope with floating-point difficulties is two-fold: first,
we employ the 80-bit Extended Precision Format provided in the x86 architecture;
second, we ensure that evaluation of any function
\begin{equation}
\begin{split}
A(x,y,z) &= \sum_{n,m,l} A_{n,m,l} x^n y^m z^l\\
&= \sum_{n,m} \biggl[ \sum_l A_{n,m,l} z^l \biggl] x^n y^m 
\equiv \sum_{n,m} \tilde{\tilde{A}}_{n,m}(z) x^n y^m \\
&= \sum_{n} \biggl[ \sum_m \tilde{\tilde{A}}_{n,m}(z)y^m \biggl] x^n
\equiv \sum_{n} \tilde{A}_{n}(y,z) x^n 
\end{split}
\end{equation}
is treated as a triply nested polynomial
\begin{equation}
\begin{split}
A(x,y,z) &= \tilde{A}_{0}(y,z) + x\bigl[\tilde{A}_{1}(y,z) + x\bigl[\tilde{A}_{2}(y,z) + \cdots \bigl]\bigl]\\
\tilde{A}_n(y,z) &= \tilde{\tilde{A}}_{n,0}(z) + y\bigl[\tilde{\tilde{A}}_{n,1}(z) + y\bigl[\tilde{\tilde{A}}_{n,2}(z) + \cdots \bigl]\bigl]\\
\tilde{\tilde{A}}_{n,m}(z) &= A_{n,m,0} + z\bigl[A_{n,m,1} +z\bigl[A_{n,m,2} +  \cdots \bigl]\bigl]
\end{split}
\end{equation}
with the floating-point operations carried out in the order dictated by Horner's rule.\cite{Borwein95}
The components of the energy gradient can be cast into the same nested form:
\begin{equation}
\begin{split}
\partial_x A(x,y,z)
&= \tilde{A}_{1}(y,z) + x\bigl[2\tilde{A}_{2}(y,z) + x\bigl[3\tilde{A}_{3}(y,z) + \cdots\bigl]\bigl]\\
\partial_y A(x,y,z)
&= \sum_n \biggl(  \tilde{\tilde{A}}_{n,1}(z) + y\bigl[2\tilde{\tilde{A}}_{n,2}(z) + \cdots \bigl]\biggr)x^n\\
\partial_z A(x,y,z)
&= \sum_{n,m} \biggl(  A_{n,m,1} + z\bigl[2A_{n,m,2} +  \cdots \bigl]\biggr)x^ny^m.
\end{split}
\end{equation}

Because the location of the energy minimum evolves smoothly as the relative coupling $g$ is varied, it is easy to
keep the computation under good numerical control as we sweep from low frustration to high. We simply
carry out the variational calculation repeatedly
in a sequence of smalls steps $g \to g + \delta g$, at each stage seeding the search with the previous step's
results.

\begin{figure}[h!]
\begin{center}
\ifthenelse {\boolean{PRBVERSION}}
{\includegraphics{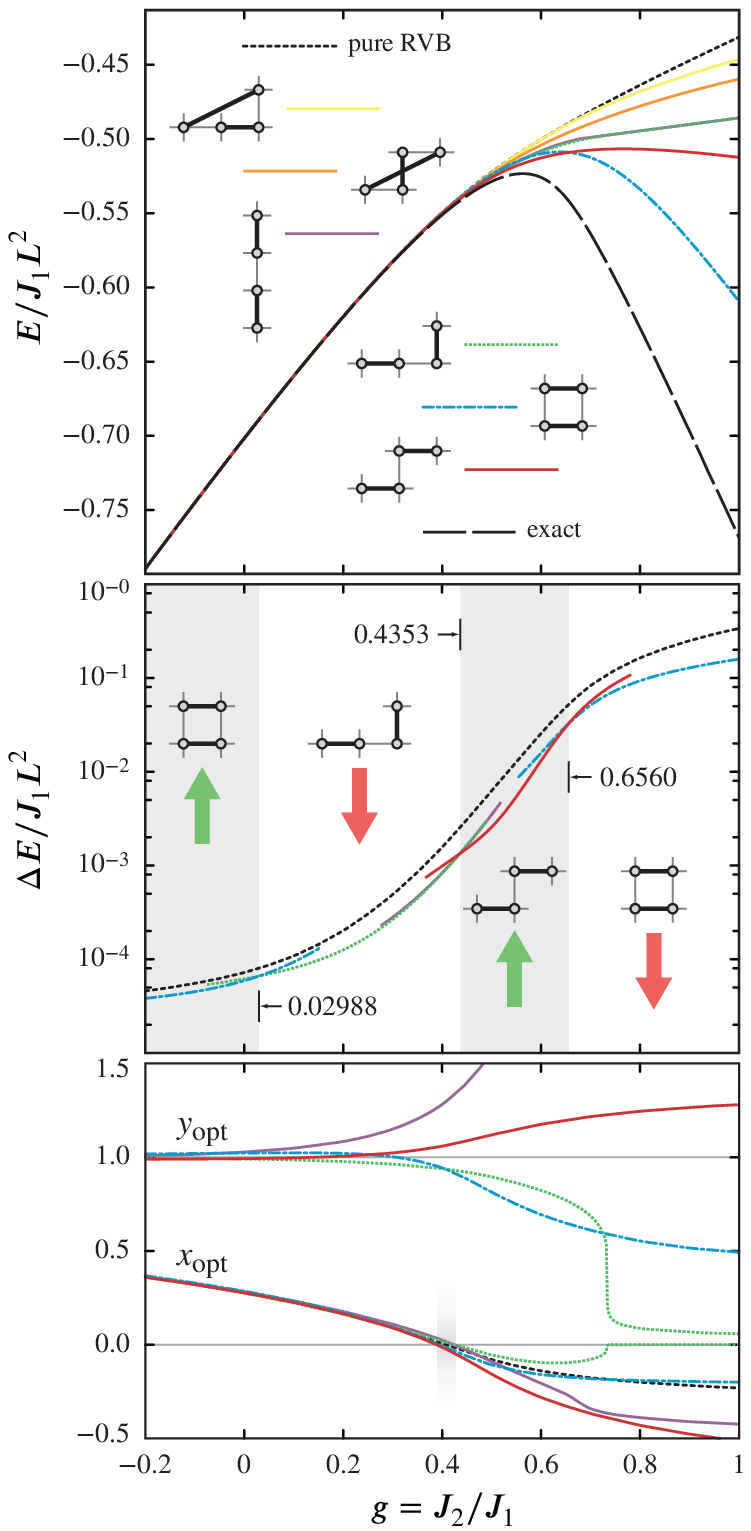}}
{\includegraphics{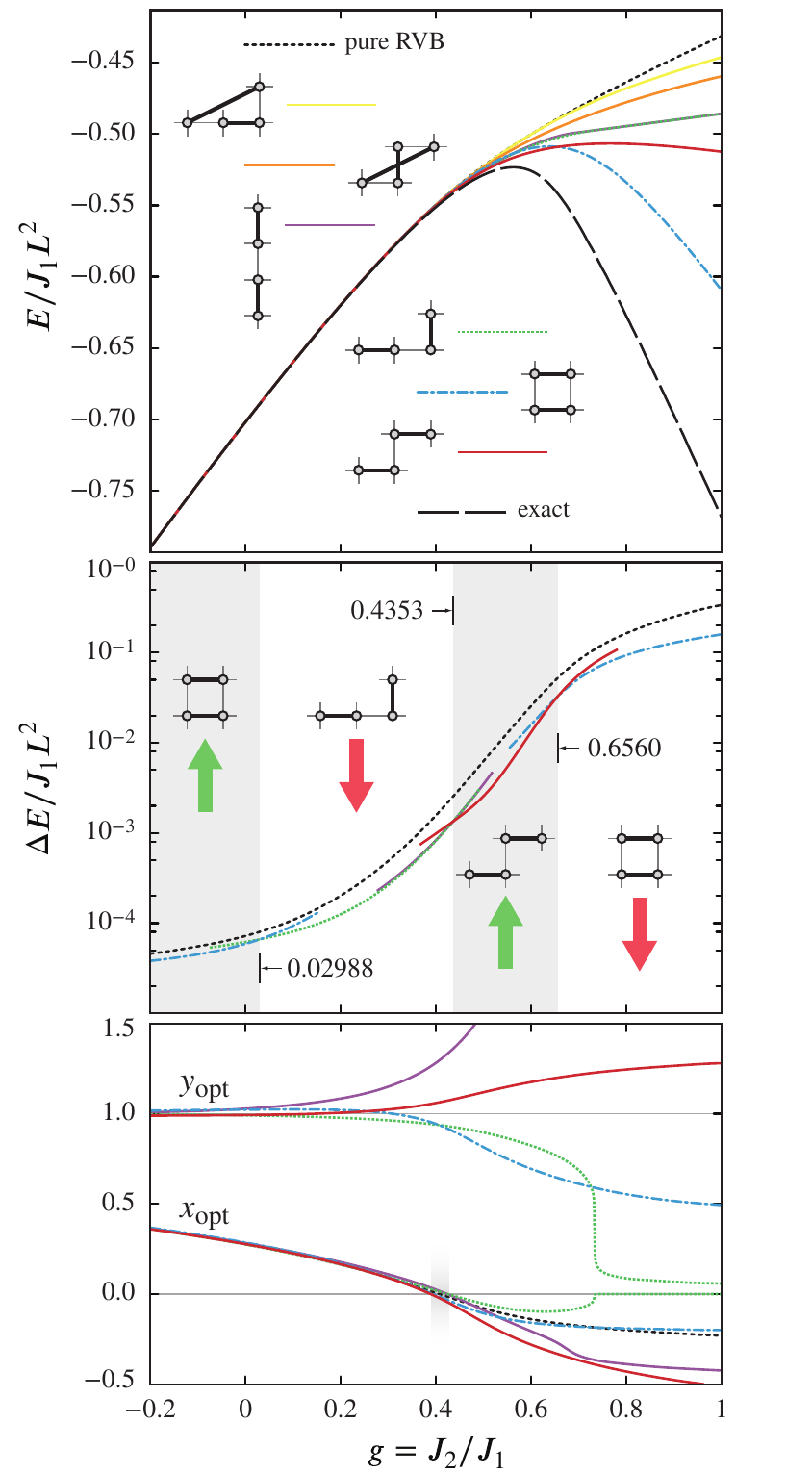}}
\end{center}
\caption{\label{FIG:energy-one} \co
(Top panel) The energy density of the RVB trial wavefunction, elaborated with at most one valence-bond-pair motif.
(Middle panel) The discrepancy between the energy density of the variational state
and that of the true ground state. The vertical stripes indicate where one motif overtakes
another as the lowest-energy state. The large up and down arrows indicate whether
the correlation factor is enhanced or suppressed.
(Bottom panel) Optimized values of the variational parameters over a large 
range of relative exchange couplings.
$x_{\text{opt}}$ vanishes
at 0.4077 (pure RVB), 
0.3935 (plaquette),
0.4238 (line),
0.3886 (staggered), and
0.4254 (herring bone).
Bond-bond correlations become pronounced above $g \approx 0.3$,
where $y_{\text{opt}}$ begins to deviate strongly from 1.
}
\end{figure}

{\it Numerical results}. The uppermost panel of Fig.~\ref{FIG:energy-one} reports the variational energy 
density of the single-motif and pure RVB (i.e., no-motif) wavefunctions. Treating the latter 
amounts to minimizing over $x=h(2,1)/h(1,0)$ alone, leaving $y=z=1$ fixed.  In the remaining cases, 
we hold $z=1$ fixed and minimize over $x$ and $y$ with $\beta(C)$ counting each of the single motifs in turn.
We find that the addition of even one bond correlation factor can substantially improve the variational energy, 
especially for large frustration.

The middle panel of Fig.~\ref{FIG:energy-one} shows the discrepancy between the energy density of the 
exact ground state and that of the best variational state determined at the current value of the relative exchange coupling.
Several distinct regimes are revealed. In the range $0.02988 < g < 0.4353$, the energy is lowest 
when the herring bone motif is slightly suppressed, its weight falling from 0.99 to 0.93 with increasing $g$.
For $0.4353 < g < 0.6560$, the best energy is given by the staggered motif, whose weight climbs from 1.08 to 1.20.
Right at the level-crossing point, $g=0.4353$, the three wavefunctions with herring bone, staggered, and line motifs are all energy degenerate.

\begin{figure}[h!]
\begin{center}
\ifthenelse {\boolean{PRBVERSION}}
{\includegraphics{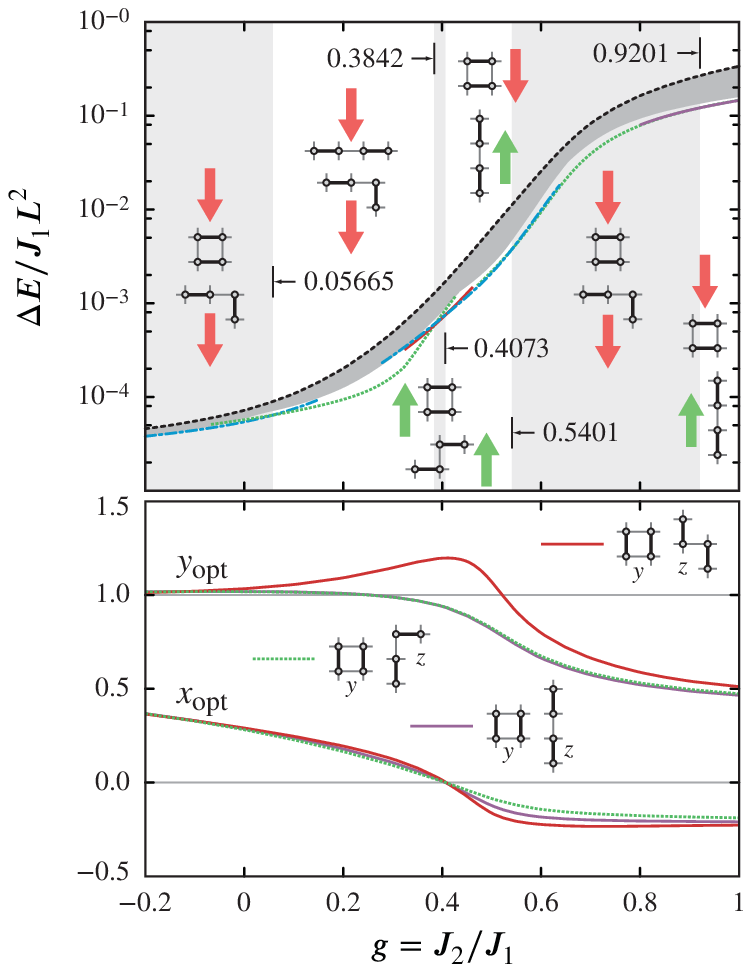}}
{\includegraphics{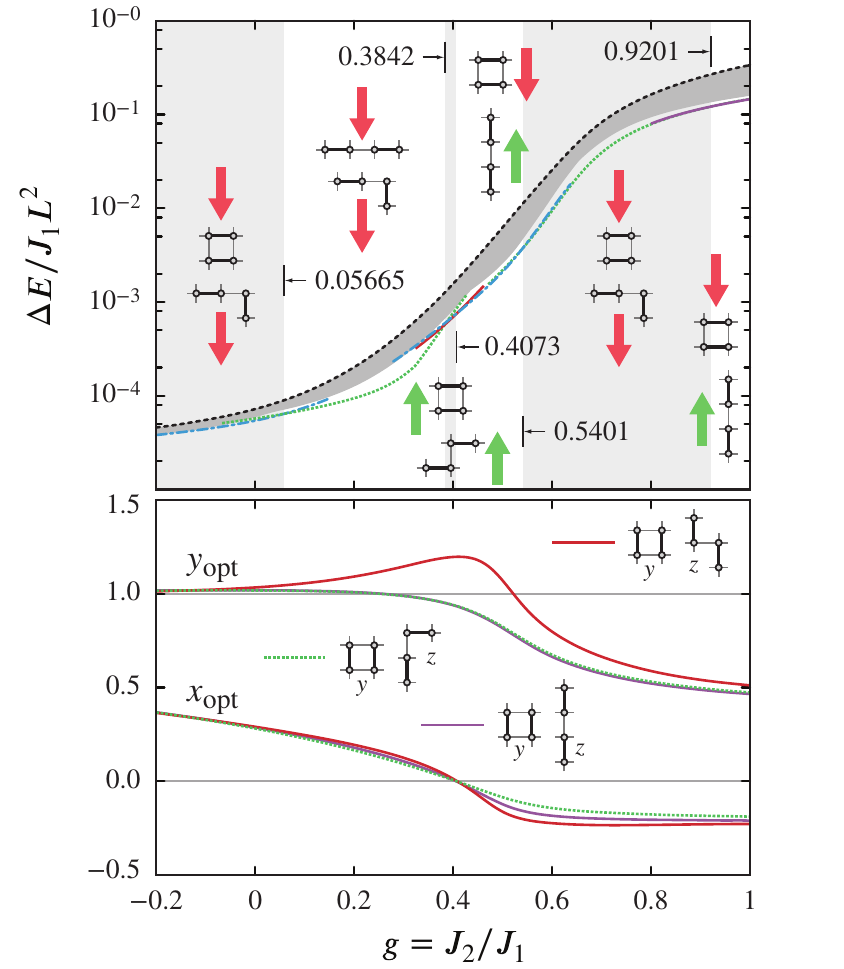}}
\end{center}
\caption{\label{FIG:energy-two} \co
(Top panel) The discrepancy between the energy density of the variational state
and that of the true ground state. Results are shown for the RVB trial wavefunction, elaborated with
two valence-bond-pair motifs and optimized over the full set of parameters $x$, $y$, and $z$.
The vertical stripes indicate where one motif overtakes another as the lowest-energy state. 
The shading underneath the pure RVB (dashed) line indicates 
the best one-motif result (i.e., the lowest value achieved in the middle panel of Fig.~\ref{FIG:energy-one}).
The large up and down arrows indicate whether the corresponding correlation factors are enhanced or suppressed.
(Bottom panel) Optimized values of the $x$ and $y$ variational parameters ($z$ is hidden) are presented
for a small subset of the two-motif wavefunctions that share a plaquette factor in common. We report
that $x_{\text{opt}}$ vanishes at
0.4085 (plaquette + staggered),
0.4073 (plaquette + herring bone), and
0.4062 (plaquette + line)
}
\end{figure}

Figure~\ref{FIG:energy-two} illustrates the further improvement to the trial state when {\it two} motifs
are taken into account. For $0.05665 < g < 0.3842$, the best variational energy is achieved
by strong suppression of the line ($0.12 < y < 0.83$) and herring bone ($0.56 < z < 0.94$) motifs.
In the thin sliver $0.3842 < g < 0.4073$, the best energy corresponds to a mild enhancement of
the plaquette ($1.19 < y < 1.20$) and staggered ($1.12 < z < 1.13$) motifs. The first of the
two ranges we have just discussed is situated in what would be the antiferromagnetic phase of the 
infinite system; the second is likely {\centred} on the critical point where the N\'eel order is extinguished. 
Note that, in contrast to the results in Fig.~\ref{FIG:energy-one}, two optimization regimes rather than one are needed
to span the magnetically disordered phase, which is believed to extend up to $g \approx 0.60$.\cite{Zhang13}
For $0.4073 < g < 0.5401$,
we find reduced plaquette ($0.75 < y < 0.93$) and increased line ($1.27 < z < 1.35$) correlations.
But deeper in the disordered phase, over a range $0.5401 < g \lesssim 0.60$ that abuts the $(\pi,0)$
antiferromagnet, the optimal state involves reductions of the 
plaquette ($0.67 < y < 0.76$) and herring bone ($0.93 < z < 0.94$) weights.

We now make some additional observations about the data in the bottom panels of 
Figs.~\ref{FIG:energy-one} and \ref{FIG:energy-two}. First, we highlight the fact that
the {\behaviour} of the optimized $x$ parameter depends only very weakly on the form 
of the variational wavefunction. Without exception, $x_{\text{opt}}$ descends monotonically
along a nearly consistent line of values and goes negative in the vicinity of $g \approx 0.41$,
which confirms our previous understanding that the Marshall rule breaks down simply by way 
of $h(2,1)$ changing sign. This appears to be a robust feature that survives the addition of 
any number of correlation factors. The second observation is that the identification of
correlation trends is not quite as clean as we had anticipated. In the bottom panel of Fig.~\ref{FIG:energy-two},
the inconsistent {\behaviour} of $y_{\text{opt}}$ shows how the plaquette correlation can be {\favour}ed
or dis{\favour}ed depending on which other bond-bond correlation factor it is combined with
(in this case, each of the staggered, herring bond, and line).
There is, however, consistency in the sense that the wavefunctions having the lowest overall energy
do agree in predicting decreased plaquette correlations throughout the magnetically
disordered phase. 

{\it Conclusions}. Our variational results indicate that the ground state of the $J_1$--$J_2$ model is 
very well approximated by a simple RVB state all the way up to $g \approx 0.3$. Beyond that level of frustration, 
explicit bond-bond correlations become increasingly important, and it is unlikely that any pure RVB 
description\cite{Lou07,Beach09,Li12,Zhang13} of the disordered phase will be reliable.

Indeed we know of several groups that have anticipated this deficiency and are considering how to 
carry out calculations with beyond-pure-RVB states, particularly in the tensor network framework\cite{Wang13}
(which, unlike stochastic sampling methods, permits evaluation of states with negative $h$
amplitudes).

Based on the small-scale (but exact) results reported here, we can offer some insights to
those attempting future large-scale, variational calculations using correlated valence bond states.
(i) The ramp and cross motifs are never competitive energetically in our scheme, so it may
be sufficient to account for correlations between short bonds only.
(ii) The sign of each configurational amplitude can be attributed to the individual bond 
amplitudes and probably even confined to just $h(2,1)$; this means that correlation factors can 
be taken to be positive definite.
(iii) In our system, the optimized values of the correlation factors are unique, but this may not be 
true generally. On larger lattices and when more than a few motifs are employed, there could be internal 
redundancy in the full set of variational parameters. (iv) The correlation weight trends do not 
always depend in an obvious way on the combination of motifs.

As for the nature of the intermediate phase, the correlations do not indicate a particular
bond order (columnar\cite{Kotov00} or plaquette,\cite{Mambrini06} say) in a straightforward 
way, but they do seem to be at odds with the columnar pattern. For instance, in the range 
$0.4073 < g < 0.5401$ we find that the weights associated with pairs of 
bonds in plaquette and line arrangements move in opposite directions (decreasing and increasing, respectively),
when we would expect them both to increase if the state were tending toward a columnar crystal.
(What this may point to instead is a resonating state with enhanced quasi-one-dimensional character,
as suggested in Refs.~\onlinecite{Beach09} and \onlinecite{Zhang13}.) Over the full range of the
disordered phase $0.4073 < g < 0.60$, the plaquette correlation is dis{\favour}ed rather significantly.

Given the level crossing that appears at $g=0.5401$, we might also entertain the possibility 
that there are multiple disordered phases\cite{Sushkov01} that intervene between
the $(\pi,\pi)$ and $(\pi,0)$ N\'eel states. For example, a reasonable scenario (one consistent with 
our results) is that the leftmost phase transition is pinned at 0.4073, where the Marshall sign rule first
breaks down; there, the $(\pi,\pi)$ antiferromagnet is superseded by a liquid phase that survives up to 0.5401, 
where a non-columnar bond crystal\cite{Saarloos00,Mambrini06} then becomes
stable; this is turn survives up to the onset of the $(\pi,0)$ antiferromagnet at 0.60. An interesting
coincidence is that $g=0.54(1)$ is the point up to which N\'eel order survives in the variational RVB
calculation with no correlation factors and all individual bond amplitudes restricted to $h > 0$.\cite{Zhang13}
This may be a clue that the state in the range $0.41 < g < 0.54$ depends in some crucial way on the
nontrivial sign structure, whereas the state in the range $0.54 < g < 0.60$ does not.

{\it {\Acknowledgements}}.  The authors offer their thanks to Anders Sandvik, Ling Wang, and 
Olexi Montunich for several useful discussions. Funding was provided by a Discovery grant 
from NSERC of Canada. KSDB also acknowledges the financial support and hospitality of the 
Department of Physics and IQIM at CalTech.


\begin{thebibliography}{99}
\bibitem{note} The full set of amplitudes irreducible by symmetry is $\{ h(a,b) \}$, where the Manhattan length $a+b$ is odd
and the coordinates sit in the ordered range $0 \le a < b \le L/2$.
\end{thebibliography}
\end{document}